\documentclass[]{IEEEtran}

\usepackage[centertags]{amsmath}
\usepackage{amssymb}
\usepackage{amsthm}
\usepackage{amsfonts, latexsym, mathrsfs,array,color,booktabs}
\usepackage{graphicx}
\usepackage{float,url}

\usepackage{color}
\usepackage[noend]{algpseudocode}
\usepackage{algorithmicx,algorithm}

\ifCLASSINFOpdf

\else

\fi

\begin{document}

\title{A Novel Feature Representation for Single-Channel Heartbeat Classification based on Adaptive Fourier Decomposition}

\author{Chunyu Tan,
        Liming Zhang,~\IEEEmembership{Member,~IEEE,}
        Hau-tieng Wu,
        and Tao Qian
\thanks{C. Tan and L. Zhang are with Faculty of Science and Technology, University of Macau, Macao, China. e-mail: (yb57416, lmzhang)@um.edu.mo.
H. Wu is with Department of Mathematics and Department of Statistical Science, Duke University, Durham, NC, USA. e-mail: hauwu@math.duke.edu. T. Qian is with Macau Institute of Systems Engineering, Macau University of Science and Technology. e-mail: tqian@must.edu.mo.}}

\markboth{Journal of \LaTeX\ Class Files,~Vol.~~, No.~~, June~2019}%
{Shell \MakeLowercase{\textit{et al.}}: Bare Demo of IEEEtran.cls for IEEE Journals}

\maketitle

\begin{abstract}
This paper proposes a novel approach for heartbeat classification from single-lead electrocardiogram (ECG) signals based on the novel adaptive Fourier decomposition (AFD). AFD is a recently developed signal processing tool that provides useful morphological features, referred to as AFD-derived instantaneous frequency (IF) features, that are different from those provided by traditional tools.
A support vector machine (SVM) classifier is trained with the AFD-derived IF features, ECG landmark features, and RR interval features.
To evaluate the performance of the trained classifier, the Association for the Advancement of Medical Instrumentation (AAMI) standard is applied to the publicly available benchmark databases, including MIT-BIH arrhythmia database and MIT-BIH supraventricular arrhythmia database, to classify heartbeats from single-lead ECG.
The overall performance in terms of sensitivities and positive predictive values is comparable to the state-of-the-art automatic heartbeat classification algorithms based on two-leads ECG.
\end{abstract}


\IEEEpeerreviewmaketitle

\section{Introduction}\label{Section:Introduction}

\IEEEPARstart{C}{ardiovascular} diseases (CVDs) are leading causes of death worldwide. According to the report of the American Heart Association (AHA) in 2018, CVDs claimed 17.9 million lives in 2015, and this number will increase to 23.6 million by 2030 \cite{JE}. Among various CVDs, arrhythmia accounts for a large proportion, and an early diagnosis of arrhythmia is of great significance to healthcare professionals \cite{Benito2015,Steinhubl2018}.

Electrocardiogram (ECG) is an effective, noninvasive, and well established diagnostic tool for arrhythmia. To achieve early diagnosis of life-threatening arrhythmias, long-term monitoring is required. Thanks to the advances of technology, this can be achieved by wearable Holters or mobile devices. As manual diagnosis of recorded long-term ECG signals is time consuming and error prone, a reliable computerized interpretation of the ECG (CIE) \cite{JS}, or at least a computer-aided automatic heartbeat annotation, has become increasingly important. Our focus in this article is a computer-aided automatic heartbeat annotation.
While there have been several commercial automatic heartbeat annotation algorithms, however, in general they show substantial rates of misdiagnosis \cite{JS}, even when they are applied to multiple-lead ECG used in the hospital.
Clearly, when there is only a single-lead ECG, the performance is deteriorated since it is more challenging to determine the delineation of fiducial points.
Wearable Holters and mobile devices that are commonly used for an early diagnosis of arrhythmia usually come with a single-lead ECG \cite{Steinhubl2018}.
Thus, it is necessary to develop an accurate automatic heartbeat classification algorithm for a single channel ECG.

A successful ECG heartbeat classification usually comprises three important procedures: preprocessing, feature extraction, and classification. In addition to the learning algorithm, feature extraction and dimension reduction are key processes that significantly affect the classification performance \cite{Houssein}.
There have been many automatic heartbeat classification algorithms, and they can be divided into two main categories based on how these three steps are carried out. The first category includes methods that are based on feature extraction and classifier training, and the second includes those based on the black-box deep learning approach \cite{PHHBN,A,X}.

\subsection{Related Works}\label{Section:RelatedWorks}

For methods in the first category, feature extraction is a critical process in automatic ECG classification analysis. An effective feature extraction method cannot only simplify computation, but also generate a superior classification performance. ECG features commonly employed for classification task are based on ECG landmarks including P, Q, R, S and T; for example, the landmark morphology \cite{ZZ,LM2} and intervals between landmarks \cite{DMR}. Researchers also consider frequency domain features based on Fourier transform \cite{GKK} and other features based on Hermite coefficients \cite{HS}, principal component analysis (PCA) \cite{YBM}, and independent component analysis (ICA) \cite{YBM}, etc.
Yet another set of features focus on the non-stationarity of the ECG into account, and researchers take the ECG time-frequency representation into; for example, short time Fourier transform (STFT) \cite{LG}, continuous wavelets transform (CWT) \cite{CWTb}, Wigner-Ville distribution \cite{AR}, and synchrosqueezing transform \cite{HFSW}.
Most studies choose support vector machine (SVM) as their classifier \cite{YBM,ZZ,HFSW}, while \cite{DMR,LM1} choose linear discriminant (LD).

The second category comprises deep learning based methods. Deep learning offers an integrated scheme that combines feature extraction and classification into one. Recently published deep learning approaches are summarized in Table \ref{Table:Comparision2}. In \cite{PHHBN}, the authors use a 34-layer deep convolutional neural network (CNN) model to detect arrhythmic heartbeats. P. Schwab et al. \cite{SCZDK2017} build a diverse ensemble of recurrent neural network (RNN) and P. Warrick et al. \cite{WH2017} use a combination of CNN with a sequence of long short-term memory (LSTM) units. In \cite{U}, four types of normal beat, congestive heart failure beat, ventricular tachyarrhythmia beat and atrial fibrillation beat are classified by combining the Lyapunov exponents algorithm with RNN. Moreover, U. Acharya et al. \cite{A} and M. Zubair et al. \cite{ZKY} develope a 9-layer and a 3-layer CNN method, respectively, whereas, \"O. Yildirim \cite{Yildirim2018} construct the deep bidirectional LSTM network-based wavelet sequences (DBLSTW-WS) network.  In \cite{MKB}, the application of the restricted Boltzmann machine (RBM) and the deep belief networks (DBN) is used for detecting ventricular and supraventricular heartbeats. However, as shown in Table \ref{Table:Comparision2}, comparison of the results across most of these deep learning approaches could not be performed due to inconsistent database and evaluation criteria. Even for the same database, the datasets used in deep learning are not the same.

There are more and more studies combining feature extraction and deep learning framework. For example, in \cite{Yildirim2018}, the DBLSTM-WS network is constructed by the wavelet-based layer and LSTM layers, and  E. \"Ubeyli \cite{U} combines Lyapunov exponents algorithm with RNN. S. Mathews et al. \cite{MKB} and G. Sannino et al. \cite{SD} take standard features, such as RR interval features, segmented morphology features and heartbeat interval features into account in their deep neural network framework.

While there have been a lot of progress and successes by methods in the second category, deep learning is regarded as black box without rigorous theoretical support. Moreover, no standard methodology exists for the construction of an optimal neural network and a lot of try-and-error with human intervention is needed.  For \cite{A}, \cite{SD} and \cite{MKB} in Table \ref{Table:Comparision2}, the authors build their network empirically by performing a wide set of trials, manually configuring the network by changing the parameters of the number of hidden layers, the activation function, the number of learning steps, and the number of neurons making up each layer.

\begin{table*}[!t]
\centering\caption{Summary of published deep learning approaches. (\# beats: number of heartbeats used, \# records: number of records used.)  \label{Table:Comparision2}}
\scalebox{0.99}{
\begin{tabular}{lccccc}
\toprule
Reference                                 & Year      &Database                  &Data size                     &Method        &Evaluation Scheme\\\midrule
U. Acharya et al. \cite{A}                &2017       &MIT-BIH arrhythmia        &452960 (\# beats)             &CNN           &intra-patient  paradigm \\
\"O. Yildirim \cite{Yildirim2018}         &2018       &MIT-BIH arrhythmia        &7326   (\# beats)             &DBLSTM-WS     &intra-patient  paradigm \\
S. Mathews et al \cite{MKB}               &2018       &MIT-BIH arrhythmia        &--     (\# beats)             &RBM, DBM      &inter-patient  paradigm  \\
G. Sannino et al \cite{SD}                &2018       &MIT-BIH arrhythmia        &4567   (\# beats)             &DNN           &inter-patient  paradigm  \\
M. Zubair et al \cite{ZKY}                &2016       &MIT-BIH arrhythmia        &100389 (\# beats)             &CNN           &patient-specific paradigm   \\
Awni Y. Hannun et al \cite{PHHBN}         &2019       & private                     &91232  (\# records)           &CNN           &F1 score        \\
P. Schwab et al  \cite{SCZDK2017}         &2017       &the physionet challenge 2017&12186 (\# records)          &RNN           &F1 score         \\
P. Warrick et al \cite{WH2017}            &2017       &the physionet challenge 2017&12186 (\# records)          &CNN+LSTM      &F1 score        \\\bottomrule
\end{tabular}}
\end{table*}

\subsection{Our contribution}

In this paper, we construct an accurate automatic heartbeat classifier with rigorous mathematical support, which belongs to the first category of methods. Our approach is motivated by viewing the limitations of existing time-frequency analysis approaches mentioned above \cite{LG,AR,ZQMD,HFSW}. Specifically, the ambiguity caused by the uncertainty principle \cite{D} in STFT \cite{LG} and CWT \cite{HFSW} is inevitable, which might mask important heartbeat information. The Wigner distribution \cite{AR} suffers from the cross-term problem \cite{F}. While synchrosqueezing transform \cite{HFSW} might help eliminate above problems, the computational complexity is not trivial and makes it unsuitable for a real-time implementation.
We thus consider a nonlinear type time-frequency analysis, the adaptive Fourier decomposition (AFD) \cite{DQG,ZQMD}, to alleviate the limitations of the above-mentioned time-frequency analysis approaches.

AFD is a novel signal processing technique with a rigorous mathematical foundation, which generalizes the traditional Fourier decomposition by taking the Blaschke decomposition theory in complex analysis into account. It decomposes a given signal by adaptively choosing its associated basis from the Takenaka-Malmquist (TM) system \cite{Takenaka,Malmquist} into a series of mono-components \cite{QW, QLM}.
By achieving maximal energy gain in each decomposition iteration, the AFD decomposes a signal into a few constitutional components called {\em mono-components}.
Those mono-components possess positive instantaneous frequencies (IF) and there is no intersection among all IFs \cite{Nahon2000,DQG}. The positive IFs effectively reflect the time-varying characteristics of signals, such as the morphology of heartbeats.
These IF-based features are used to design an automatic heartbeat classification algorithm. Specifically, we train a SVM classifier by taking these IF-based features and some basic landmark features into account. We mention that AFD can be applied to other biomedical signal analysis; for example, it has been successfully applied to the ECG signal compression problem \cite{MZD, TZW}.
To validate the proposed automatic heartbeat classifier, an inter-patient cross validation paradigm is applied. The Association for Advancement of Medical Instrumentation (AAMI) standard \cite{AAMI1987,Standard2008} along with the benchmark MIT-BIH arrhythmia database from Physionet \cite{PhysioBank} are used for a comprehensive comparison with other ECG classification methods. Moreover, the combination of MIT-BIH arrhythmia database and MIT-BIH supraventricular arrhythmia database from Physionet \cite{PhysioBank} is considered to train our final classifier.

\subsection{Organization}
The rest of the paper is organized as follows: the mathematical background of the proposed automatic feature representation and extraction are elaborated in Section \ref{Section:MathFoundation};  Section \ref{Section:Methods} details the proposed method adopted for the classification problem; the results of the classification performance are presented in Section \ref{Section:Results} and comparisons with state-of-the-art algorithms are discussed in Section \ref{Section:Dicussion}, with the conclusion provided in Section \ref{Section:Conclusion}.

\begin{figure*}[!htbp]
  \centering
  \includegraphics[width=.99\linewidth]{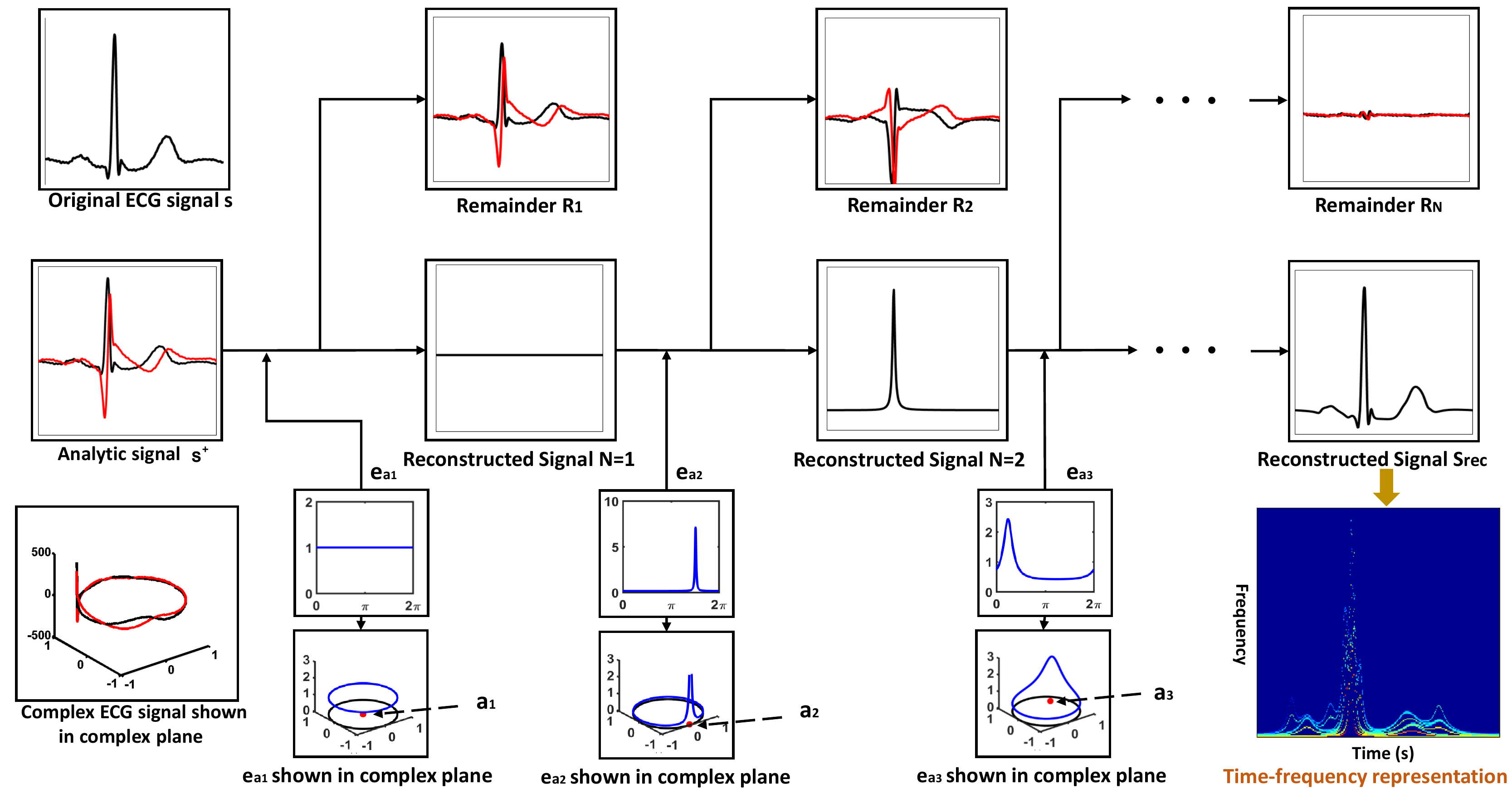}
  \caption{The flowchart of  AFD with a real ECG signal. The black and red curves denote the real and imaginary parts of a complex signal, respectively. Note that $a_l$, where $l=1,2,3$, are shown as red dots in those plots showing $e_{a_l}$, where $l=1,2,3$. In the time-frequency representation shown in the right bottom subplot, note that the instantaneous frequencies of all decomposed components are shown together.}
  \label{fig:AFDflowchart}
\end{figure*}

\section{Mathematical Foundation}\label{Section:MathFoundation}

\begin{figure*}[!htbp]
  \centering
  \includegraphics[width=.99\linewidth]{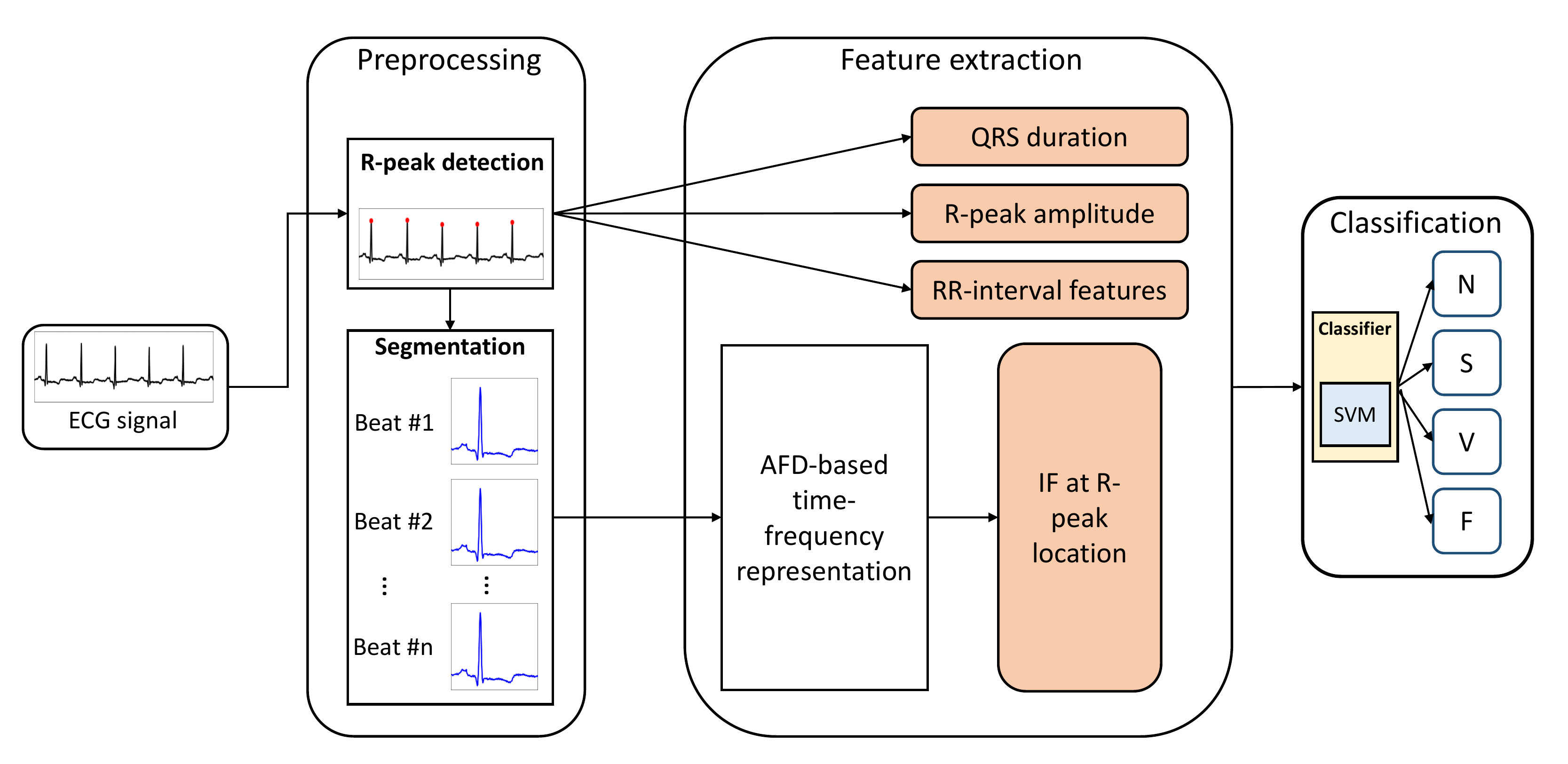}
  \caption{Overview of the proposed automatic heartbeat classification algorithm.}
  \label{fig:flowchart}
\end{figure*}

\subsection{AFD-based time-frequency representation}
Take a set of complex numbers $\mathcal{A}:=\{a_{n}\}_{n=1}^\infty\subset \mathbb{D}$, where $\mathbb D\subset \mathbb{C}$ is the unit disc and $a_n$ may repeat. The TM system \cite{Takenaka,Malmquist} associated with $\mathcal{A}$ is defined as $\{B_{n}\}$, where
\begin{equation}\label{000}
  B_{n}(z)=\frac{\sqrt{1-|a_{n}|^{2}}}{1-\overline{a_{n}}z}\prod_{k=1}^{n-1}\frac{z-a_{k}}{1-\overline{a_k}z}\,
\end{equation}
and $n\in \mathbb{N}$.
$B_{n}$ is referred to as a \em modified Blaschke product\em, while $\prod_{k=1}^{n-1}\frac{z-a_{k}}{1-\overline{a_k}z}$ is the usual Blaschke product. In the literature, the term $\frac{\sqrt{1-|a_{n}|^{2}}}{1-\overline{a_n}z}$ in \eqref{000} is called the {\em evaluator} at $a_n$. It can be shown that the TM system is an orthogonal system.
We call a TM system {\em adaptive} if we select $a_{n}$ according to the input signal.

Take a real-valued function $s\in L^{2}(\partial\mathbb{D})$, where $\partial\mathbb{D}$ is the unit circle. The associated analytic signal, $s^{+}$, is defined as
\begin{equation}\label{1}
  s^{+}=\frac{1}{2}\big(s+iHs+c_{0}\big),
\end{equation}
where $c_{0}$ is the $0$-th Fourier coefficient, and $H$ is the Hilbert transform.

For the analytic signal $s^{+}$, AFD conducts fast converging expansion of $s^{+}$ in orthogonal terms of the form
\begin{equation}\label{0004}
s^{+}=\sum_{n=1}^{\infty}c_{n}B_{n},
\end{equation}
under a selection of the parameters $a_{1},\ldots,a_{n},\ldots$, by maximal selection principle \cite{QW}, where $c_{n}:=\left\langle s^{+},B_{n}\right\rangle$ is the n-th coefficient of $B_n$.

By denoting
\begin{equation}\label{0004}
s^{+}_{rec}=\sum_{n=1}^{N}c_{n}B_{n},
\end{equation}
where $N\in \mathbb{N}$ is called the {\em decomposition level} and $s^+_{rec}$ is called the {\em AFD-approximation of degree $N$}.
Then, we have
\begin{equation}\label{004}
s^{+}=s^{+}_{rec}+R_{N}\,.
\end{equation}
It was proved in \cite{QW} that $s^{+}_{rec}$ converges to $s^{+}$ in the $H^{2}$ convergence sense; that is, $\|R_{N}\|_{H^{2}}\rightarrow0$ as $N\rightarrow\infty$. See Fig. \ref{fig:AFDflowchart} for a detailed illustration of the AFD for the ECG signal.

By (\ref{1}), clearly
\begin{equation}\label{5}
  s_{rec}=2\Re s_{rec}^{+}-c_{0},
\end{equation}
where $\Re$ means taking the real part. $s_{rec}$ is the approximation of $s$.
We mention that theoretically $\Re s^{+}$ might be deviated from $s$. Thus, an accurate approximation of $s^+$ by $s^{+}_{rec}$ does not mean an accurate approximation of $s$ by $\Re s^{+}_{rec}$. When $s$ is an ECG signal we have interest, empirically we see that $\Re s^{+}$ approximates $s$ accurately. See Figure \ref{fig:VFIF} for examples when $N=10$. On the other hand, since our purpose is feature extraction, an accurate approximation of $s$ is not critical for our application.

To visualize how a given real-valued signal $s$ oscillates, it is natural to consider a time-frequency representation. There are several different approaches to generate time-frequency representations of a given signal, ranging from linear to non-linear types \cite{D,F}. For the AFD, we consider the following approach to construct the time-frequency representation.
Denote $c_{n}B_{n}$, $n=1,\ldots,N$ to be the $n$-th level AFD of $s$. If $c_{n}B_{n}(e^{it})=\rho_{n}(t)e^{i\theta_{n}(t)}$, where $t\in [0,2\pi)$ is the time, the {\em transient time-frequency representation} of $s$ proposed in \cite{DQG,ZQMD} is then defined as
\begin{equation}\label{b3}
  R_s(t,\zeta)=\sum_{n=1}^N\rho_n^{2}(t)\delta(\theta_n'(t)),
\end{equation}
where $\zeta>0$ is the frequency, and $\delta$ is the distributional Dirac function \cite{DQG}. Note that $R_s$ can be numerically plotted as an image for the purpose of visualization as shown in Fig. \ref{fig:AFDflowchart}.

\subsection{The instantaneous frequency (IF) feature}

A function $s(e^{t})=\rho(t)e^{i\theta(t)}\in L^{2}(\partial\mathbb{D})$ is called a {\em mono-component} if $\rho\geq0$ and $\theta'\geq0$ a.e. \cite[Definition 1.1]{Q3}; that is, it has well defined non-negative analytic phase derivatives.
The non-negative analytic phase derivative $\theta'$ is the IF of $s$.

\begin{figure}[!htbp]
  \centering
  \includegraphics[width=.99\linewidth]{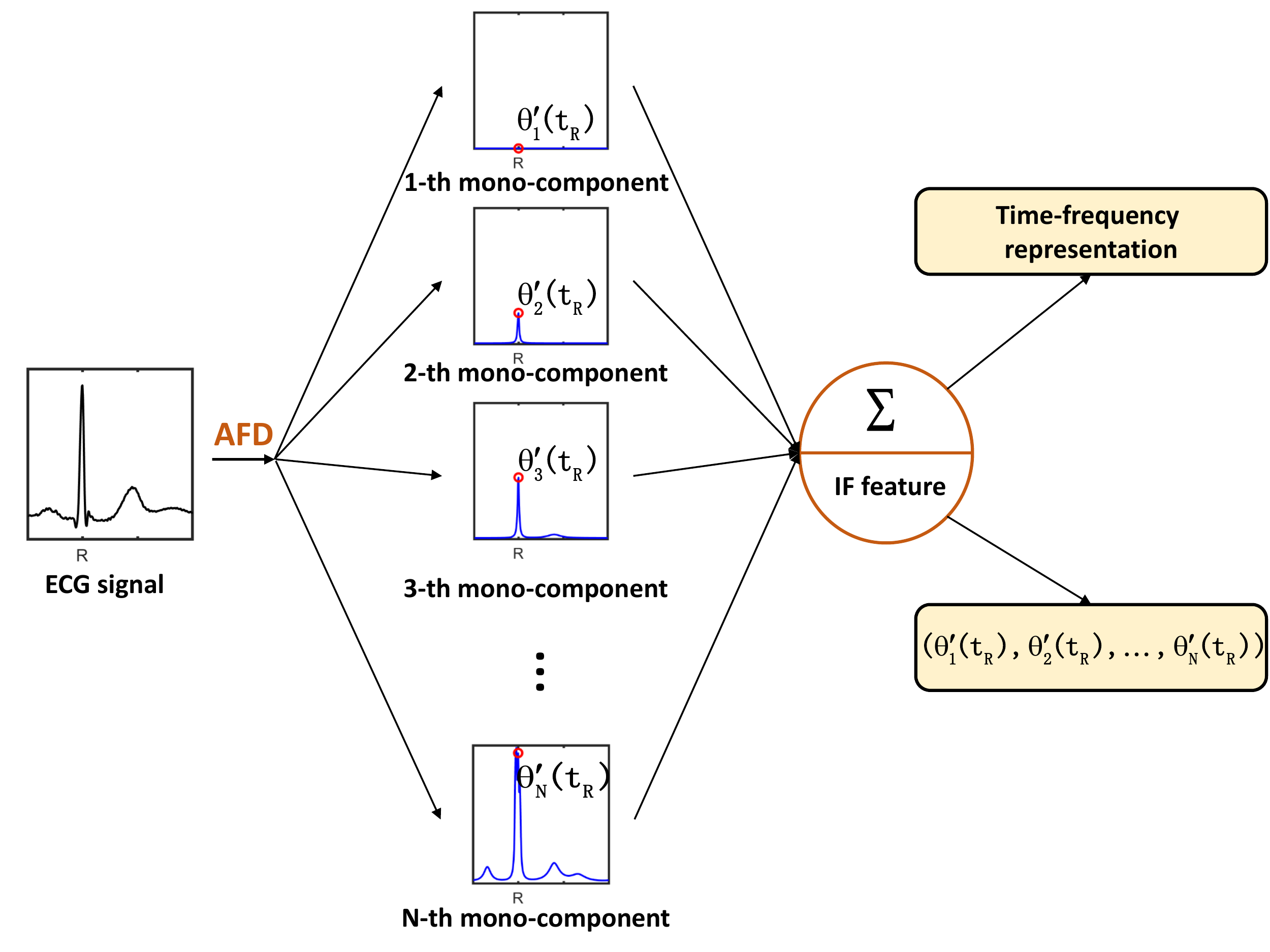}
  \caption{Process of extracting IF feature vectors.}
  \label{fig:ITfeature}
\end{figure}

According to (\ref{0004}) and (\ref{b3}), if $c_{n}B_{n}(e^{it})=\rho_{n}(t)e^{i\theta_{n}(t)}$, by a direct calculation, we have
\begin{align}\label{06}
   \theta'_{n}(t)=&\frac{|a_{n}|\cos(t-\theta_{a_{n}})-|a_{n}|^{2}}{1-2|a_{n}|\cos(t-\theta_{a_{n}})+|a_{n}|^{2}}\\ \nonumber
                  &+\sum_{l=1}^{n-1}\frac{1-|a_{l}|^{2}}{1-|a_{l}|\cos(t-\theta_{a_{l}})+|a_{l}|^{2}},
\end{align}
where $\theta_{a_n}=|a_{n}|e^{i\theta_{a_{n}}}$.
We view the IF function $\theta'_{n}$ as a feature of the given signal.
In our application, if $t_R$ represents the location of a R-peak,  we call $\theta'_{n}(t_R)$ the {\em $n$-th R-peak IF feature} of that heartbeat. For the decomposition level $N$, we call the vector $(\theta'_{1}(t_R),\theta'_{2}(t_R),\ldots,\theta'_{N}(t_R))$ the {\em R-peak IF feature vector} of a given heart beat. The process of extracting the R-peak IF feature vector is described in Fig. \ref{fig:ITfeature}.

\begin{table*}[!t]
\centering
\caption{Relationship between heartbeat types in the MIT-BIH arrhythmias database and heartbeat classes in the AAMI standard as well as the number of heartbeats in the training and testing sets. (N: normal; L: left bundle branch block; R: right bundle branch block; A: atrial premature; a: aberrated atrial premature; J: nodal (junctional) premature; S: supraventricular premature or ectopic; V: premature ventricular contraction; F: fusion of ventricular and normal; e: atrial escape; j: nodal (junctional) escape; E: ventricular escape; Q: unclassifiable beat.) \label{Table:AAMIStandard}}
\scalebox{0.99}{
\begin{tabular}{l|ccccccccccccc|c}
\toprule
Heartbeat types           &N      &L    &R      &A    &a    &J    &S   &V      &F    &e    &j   &E    &Q &   \\
Heartbeat classes         &N      &N    &N      &S    &S    &S    &S   &V      &F    &N    &N   &V    &Q &Total\\\midrule
All records               &74173  &8038 &7233   &2535 &149  &84   &2   &6730   &801  &14   &223 &106  &15&100103\\
DS1                       &37917  &3933 &3764   &807  &99   &33   &2   &3648   &423  &14   &12  &105  &8 &50765\\
DS2                       &36256  &4105 &3469   &1728 &50   &51   &0   &3082   &378  &0    &211 &1    &7 &49338\\\bottomrule
\end{tabular}}
\end{table*}

\subsection{A related approach}

The applied AFD is directly related to the {\em Blaschke decomposition (BKD)} algorithm considered in \cite{CS2015,Nahon2000,CSW2017}, and a side-by-side comparison would highlight the essence of the proposed algorithm. The main difference between AFD and BDK is the usage of the TM system. In BDK, instead of applying the maximal selection principle, it is $0$ that is considered to produce zeros inside the unit disk, which leads to the next decomposition. Also, the BKD is applied at each decomposition step. As is developed in \cite{Nahon2000}, the decomposed components all have positive IF.
After its appearance in \cite{Nahon2000}, several studies have attempted to explore the properties of BKD. For example, R. Coifman et al. \cite{CS2015} investigated the unwinding property and the convergence behavior of BKD; in \cite{Saito,CSW2017,LNX2018}, BKD is explored from an algorithmic perspective guided by theoretical development; in \cite{CP2017,SW2018}, the fundamental root distribution properties associated with BKD were studied. These developed theories are all directly related to the AFD.
Finally, while we do not yet have theoretical support, numerically, overall, the behavior of the AFD and BKD is ``similar'' with a delicate difference.

\section{Materials and Methods}\label{Section:Methods}

\subsection{Dataset}

The well-known MIT-BIH arrhythmia database\footnote{\url{http://www. physionet.org/physiobank/database/mitdbl}} from Physionet \cite{PhysioBank} is used for training and testing purposes.
The MIT-BIH arrhythmia database contains 48 recordings of approximately 30 minutes, each of which has two leads, lead A and lead B.
In 45 recordings, lead A is MLII (modified lead II), and lead B is mainly V1, but sometimes V2, V4 or V5; in the remaining 3 recordings, lead A is V5 and lead B is V2 or II. The signals are sampled at $360$ Hz. We use lead A in this work for the automatic heartbeat classification.
Not all recordings are the same in virtue of the arrhythmia and physical limitation of the subjects.
Twenty-three recordings serve as representative samples of routine clinical recordings, and 25 recordings contain complex ventricular, junctional and supraventricular arrhythmias.
The database comes with experts' annotation. The annotations follow the AAMI standard \cite{AAMI1987,Standard2008}, which further categorizes beat types into different classes, as is shown in Table \ref{Table:AAMIStandard}.
Specifically, the N class contains beats originating in the sinus node (normal and bundle branch block beat types); the S class contains supraventricular ectopic beats; the V class contains ventricular ectopic  beats; the F class contains beats that result from fusing normal and ventricular ectopic beats; the Q class contains unknown beats, including paced beats. In this work, the Q class is discarded according to the recommended practice since it is marginally represented in the database.
The provided experts' annotations are used as the standard to evaluate the performance of classification result.

In addition to the MIT-BIH arrhythmia database, the MIT-BIH supraventricular database\footnote{\url{http://www. physionet.org/physiobank/database/svdb}} is used to alleviate the imbalanced classes issue when working with the MIT-BIH arrhythmia database. Specifically, there are limited beats of type S in the MIT-BIH arrhythmia database.
The MIT-BIH supraventricular database consists of 78 recordings of approximately 30 minutes sampled at 128 Hz, each of which has two leads. %
The ECG recordings are resampled to 360 Hz to match the sampling rate of the MIT-BIH arrhythmia database.
Based on the AAMI standard and the provided labels, there are 12,148 beats of class S, which are used to train the automatic heartbeat classifier.

\subsection{Heart beat classification algorithm}

We assume that the input ECG signal is sampled at 360 Hz. The proposed automatic heartbeat classification algorithm consists of three steps, namely, preprocessing, feature extraction and classifier construction.
In the first step, the raw ECG signals are divided into heartbeat segments following the standard R peak detection algorithm. Next, the AFD is applied to generate the IF feature vector for each heartbeat segment. A set of commonly applied landmark features, including the QRS duration, R-peak amplitude and RR-interval, are also derived. Finally, a SVM classifier is trained for the purpose of classification.
See Fig. \ref{fig:flowchart} for an illustration of the proposed algorithm.

\subsubsection{Preprocessing}

The preprocessing stage consists of R peak detection and heartbeat segmentation.
Since the R peak detection is not the focus of this work, the R peak annotations provided in the MIT-BIH arrhythmia database are used for the heartbeat segmentation.
For each detected R peak, 100 sampling points before the R peak location and 200 sampling points after it are selected to construct an associated heartbeat segmentation. Considering that the sample rate is 360Hz, a heartbeat segmentation is approximately 0.83s.
Note that due to the fundamental property of AFD, the denoising step and the detrending step commonly considered in the preprocessing process of most methods are not needed.

\begin{figure*}[!htbp]
  \centering
  \includegraphics[width=.99\linewidth]{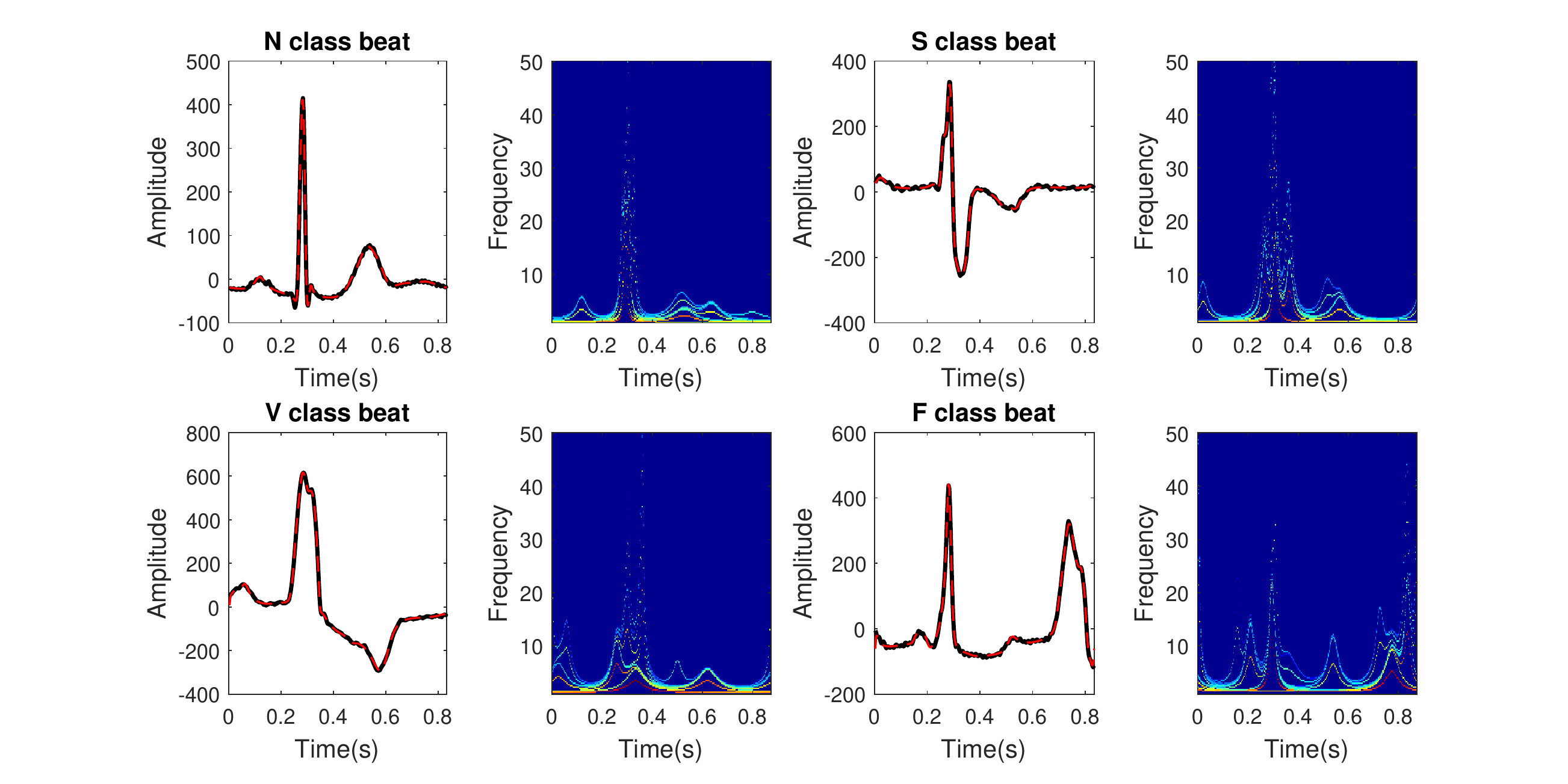}
  \caption{Results of AFD-approximation and the associated time-frequency representations of four different heartbeat classes. The decomposition level is $N=10$. Black lines represent the original heart beats and red dotted lines represent the respective approximation.}
  \label{fig:VFIF}
\end{figure*}

\subsubsection{Feature extraction}

\begin{figure*}[!htbp]
  \centering
  \includegraphics[width=.6\linewidth]{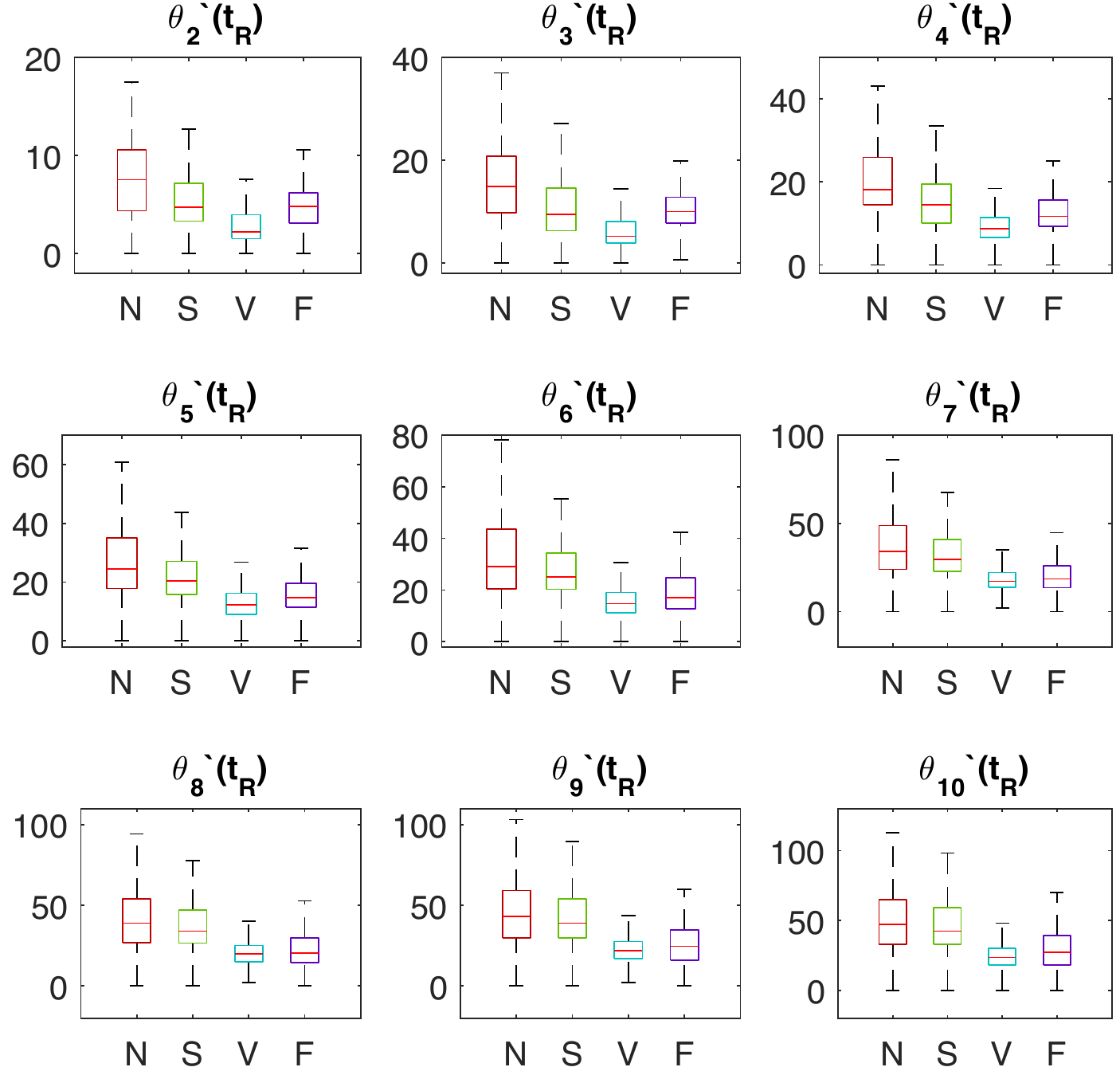}
  \caption{Boxplots of the distribution of IFs of nine mono-components obtained by AFD at R locations for the main beat types presented in the MIT-BIH database. Note that we set all $a_{1}=0$ for N=1 and omit it. N: N class; S: S class; V: V class; F: F class.}
  \label{fig:IF_boxplot}
\end{figure*}

For each heartbeat segment, the AFD is applied with the decomposition level $N=10$. $N=10$ is chosen so that the decomposed mono-components well recover the heartbeat segment.
Thus, each heartbeat segment is decomposed into nine mono-components since the first mono-component is trivial by the choice of $a_{1}=0$. The IFs of these nine mono-components at the R peak locations are chosen as features of each heartbeat segment; that is, the R-peak IF feature vector is $(\theta'_{2}(t_R),\theta'_{3}(t_R),\ldots,\theta'_{10}(t_R))\in \mathbb{R}^9$.
The distributions of nine R-peak IF features of the N, S, V, and F class are graphically represented in Fig. \ref{fig:IF_boxplot}. Visually we see that the IF feature vectors are discriminative representations of the heartbeat types.
Moreover, we consider another five features to capture the morphological information related to the P wave. In general, the PR interval ranges from 0.12 to 0.2 seconds but not fixed. Thus, for each heartbeat segment, the IFs at the 50th samples before the R peak location of the first five mono-components are considered; that is, $(\theta'_{2}(t_R-50),\theta'_{3}(t_R-50),\ldots,\theta'_{5}(t_R-50),\theta'_{6}(t_R-50))\in \mathbb{R}^5$, which we call the {\em P-wave IF feature vector}.

In addition, landmark features and dynamic features that have been clinically studied with stipulated diagnostic standards \cite{ECG} were also chosen. In total, we have $19$ features for each heartbeat segment, which are listed in Table \ref{Table:Features}.

\begin{table}[!h]
\centering\caption{The feature set in this study. \label{Table:Features}}
\scalebox{0.99}{
\begin{tabular}{l|l}
\toprule
Features           &Description         \\\midrule\midrule
R-peak IF feature vector & AFD-derived IFs at R peaks.\\\midrule
P-wave IF feature vector & AFD-derived IFs at P waves.\\\midrule
QRS duration       &Duration of the QRS complex.      \\\midrule
R-peak amplitude   &Amplitude of the R point location.       \\\midrule
Pre-RR intervals   &time difference between current \\
                   &and previous beat at R-peak.         \\
Post-RR intervals  &time difference between current \\
                   &and the next beat at R-peak.  \\
Local-RR intervals &Average R peak to R peak \\
                   &interval over 10 beats.\\\bottomrule
\end{tabular}}
\end{table}

\subsubsection{Classifier}

We consider the widely applied classifier with solid theoretical foundation, kernel SVM \cite{VV}, to establish the heartbeat classification model.
The radial basis function (RBF) kernel is considered in this work, which is defined as
\begin{equation}\label{ab1}
  K(x_{i},x_{j})=\exp\Big(-\frac{\|x_{i}-x_{j}\|^{2}}{2\sigma^{2}}\Big)\,,
\end{equation}
where $\sigma>0$ is the bandwidth, and $x_{i}$, $x_{j}$ are the feature vectors of the $i$th and $j$th heartbeat segment, respectively.
Numerically, the libsvm library \cite{LIBSVM} is applied to implement the kernel SVM. Libsvm possesses a flexible interface and supports multi-class classification by one-versus-one approach.
Since the training data from the MIT-BIH arrhythmia database is imbalanced, the weighted SVM classifier is trained. Specifically, we add each class a weight to penalize the class according to its prevalence to relieve the problem generated by the imbalance of the training dataset.

\subsection{Inter-patient cross validation}

To evaluate the performance of the proposed algorithm, cross validation (CV) is performed.
Numerous works report that automatic heartbeat classification can be categorized into two types of CV, namely the intra-patient CV or inter-patient CV. When the training and testing sets contain heartbeats from the same subjects, it is called the intra-patient CV, otherwise it is known as the inter-patient CV \cite{DMR,HFSW,phenotype}. Intra-patient CV has been widely adopted in most works and achieves optimistic results, but it is not suitable for real situations. Following the state-of-the-art approaches \cite{DMR,phenotype,HFSW}, the inter-patient CV is performed in this study.

The whole MIT-BIH arrhythmia database is divided into the training set (DS1) and the test set (DS2) \cite{DMR}, so that DS1 and DS2 are from recordings from different patients. Note that the total number of beats from DS1 and DS2 in this study are slightly different from \cite{DMR}.  The first 10 beats from each recording are dropped, because the Local-RR intervals are calculated by the average of the previous 10 beats. This value starts to be calculated from the 11th beats. We also drop the last beat, as there is no Post-RR interval for this beat. At last, we have total 100103 beats in our study (comparing 100731 beats in [11]).

\begin{figure}[!htbp]
  \centering
  \includegraphics[width=.99\linewidth]{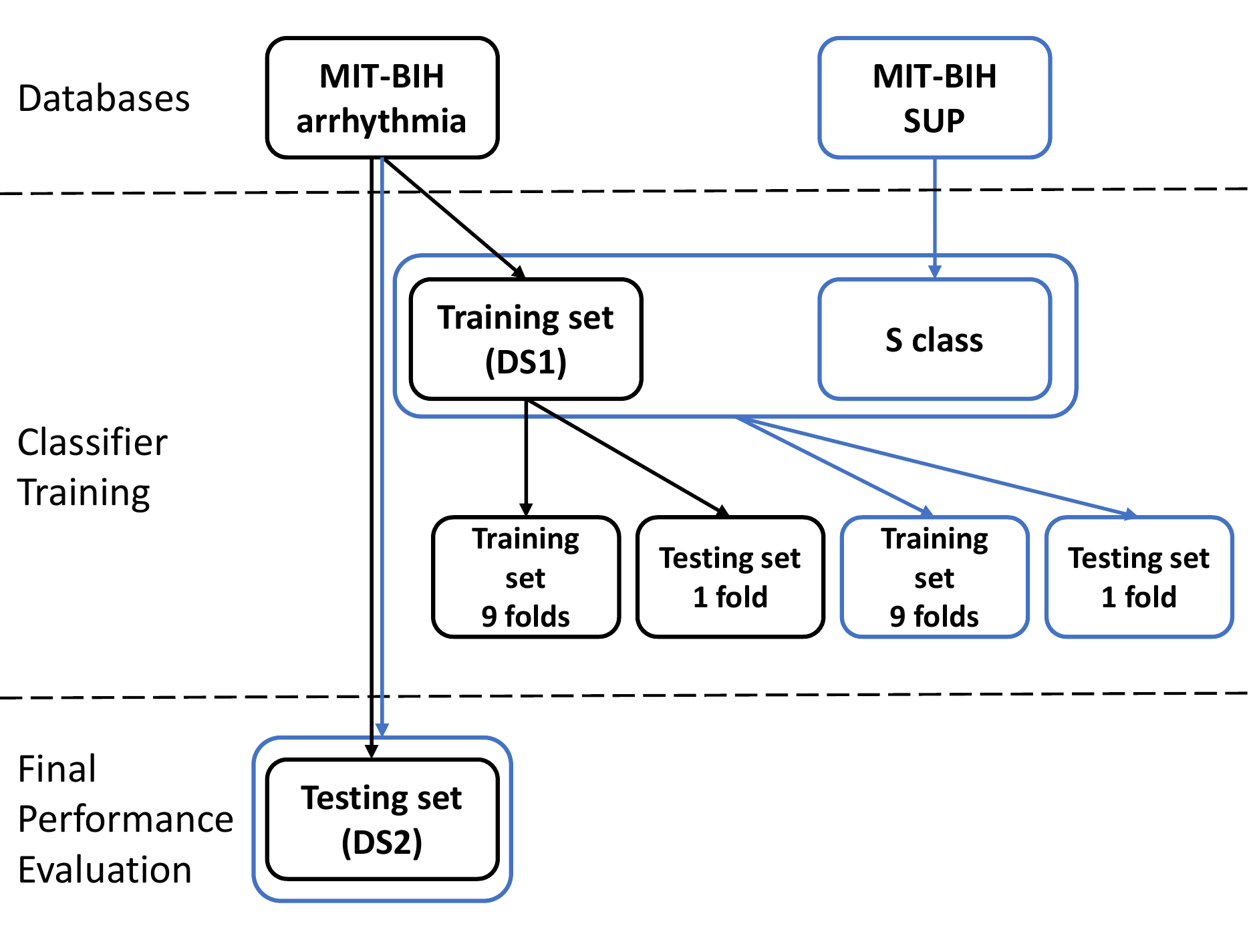}
  \caption{Visualization process for the performance assessment of the classifier.}
  \label{fig:Visualization}
\end{figure}

We consider the following two models.
To get the first model, the classifier is trained on all features listed in Table \ref{Table:Features} and annotations from DS1. The optimal parameters for the weighted kernel SVM are determined by the grid optimization method by applying the 10-fold CV.
The established SVM classifier is then validated on DS2.
To get the second model, the classifier is trained all features listed in Table \ref{Table:Features} and annotations from DS1 and the heartbeats in the S class from the MIT-BIH supraventricular database, and then assessed on DS2.
See Fig. \ref{fig:Visualization} for an illustration of the two CV schemes, in which the black block diagram shows the process of the first stage and the second stage is presented by the blue block diagram.

\subsection{Performance Evaluation}
The confusion matrix shows a detailed distribution of the classification results achieved by a classifier, hence an evaluation of the performance of a classifier. When $m$ labels are involved in the classification, the confusion matrix is a $m\times m$ square matrix.
Denote $n_{kl}$ to be the $(k,l)$-th entry of the confusion matrix. The sensitivity (Se) and positive predictivity (+P) for the k-th class is defined as
\begin{equation}\label{10}
  Se_{k}=\frac{n_{kk}}{\sum_{l=1}^{N}n_{kl}}
\mbox{ and }
  +P_{k}=\frac{n_{kk}}{\sum_{k=1}^{N}n_{kl}}
\end{equation}
respectively. The overall accuracy (Acc) is denoted as
\begin{equation}\label{09}
  Acc=\frac{\sum_{k=1}^{N}n_{kk}}{\sum_{k=1}^{N}\sum_{l=1}^{N}n_{kl}}.
\end{equation}

\section{Results}\label{Section:Results}
Regarding the AFD-based time-frequency representation of ECG signals, Fig. \ref{fig:VFIF} shows the results for four heartbeat segments from different classes, including the AFD-approximation of degree $10$ and their respective time-frequency representations.

For the first model, the optimal parameters for the kernel SVM classifier determined from DS1 are $C=2.8$, $\sigma=0.0005$, $\omega_{1}=0.40$, $\omega_{2}=39$, $\omega_{3}=2.9$ and $\omega_{4}=1.71$, where $\omega_1$, $\omega_2$, $\omega_3$ and $\omega_4$ are weights for classes N, S, V, F respectively. The confusion matrix is shown in Table \ref{Table:Confusion Matrix1}. This classifier assessed on DS2 achieves a 84.92\% overall accuracy and sensitivity of 85.48\%, 78.08\%, 81.19\% and 83.33\% for the F, N, S, and V class respectively.
As shown in Table \ref{Table:Confusion Matrix1}, the performance of the S class is slightly poor and improving it is challenging, mainly due to the imbalanced dataset issue.

For the second model,
the optimal parameters were found to be $C=3$, $\sigma=0.0006$, $\omega_{1}=0.42$, $\omega_{2}=36$, $\omega_{3}=2.5$ and $\omega_{4}=1.79$. The confusion matrix is shown in Table \ref{Table:Confusion Matrix2} and the final performance of the SVM classifier testing on DS2 had an 85.02\% overall accuracy. The detailed Se and +P of each class are shown in Table \ref{Table:Comparision}.
Note that with the help of the MIT-BIH supraventricular database, the result of the S class is improved, and hence the overall performance.

\begin{table}[!h]
\centering\caption{Confusion Matrix 1. \label{Table:Confusion Matrix1}}
\scalebox{0.99}{
\begin{tabular}{c|cccccc}
\toprule
                 &\multicolumn{5}{|c}{Predicted}              \\\midrule
                 &              & N              &S              &V            & F  \\\midrule
Reference        &N             & 37647          &3625           &260          & 2509           \\
                 &S             & 339            &1428            &61          & 1              \\
                 &V             & 61             &415            &2503         & 104             \\
                 &F             & 36             &5              &22           & 315              \\\bottomrule
\end{tabular}}
\end{table}

\begin{table}[!htbp]
\centering\caption{Confusion Matrix2. \label{Table:Confusion Matrix2}}
\scalebox{0.99}{
\begin{tabular}{c|cccccc}
\toprule
                 &\multicolumn{5}{|c}{Predicted}\\\midrule
                 &                &N             & S             &V            & F  \\\midrule
Reference        &N             & 37681               &3555             & 231              & 2574               \\
                 &S             & 299               &1470             & 58              & 2              \\
                 &V             & 67               &433             & 2477              & 106             \\
                 &F             & 38               &7             & 20              & 313              \\\bottomrule
\end{tabular}}
\end{table}

\section{Discussion}\label{Section:Dicussion}

We propose a novel signal feature representation approach, which is applied to extract time-frequency features from the ECG signal for the automatic heartbeat classification. The proposed algorithm is trained and validated on publicly available databases with the inter-patient CV.

\begin{table*}[!t]
\centering\caption{Comparison of the proposed method with previous works. (\# leads: number of ECG leads used; \# features: number of features extracted.) \label{Table:Comparision}}
\scalebox{0.99}{
\begin{tabular}{lccccccccccc}
\toprule
                                  &&        &\multicolumn{2}{c}{N} &\multicolumn{2}{c}{S}      &\multicolumn{2}{c}{V}     &\multicolumn{2}{c}{F}       &Tot.\\\midrule
Method                           &\# leads &\#features                 &Se    &+P    &Se     &+P    &Se    &+P    &Se     &+P     &Acc  \\\midrule
P. De Chazal et al \cite{DMR}    &2  &52             &86.9  &99.2  &75.9   &38.5  &77.7  &81.9  &89.43  &0.08   &81.9\\
M. Llamedo et al \cite{LM1}    &2    &39           &77.55 &99.47  &76.46   &41.34  &82.94  &87.97  &95.36  &4.23   &78.0\\
C. Ye et al \cite{YBM}    & 2   & 132          &88.5  &97.5  &60.8   &52.3  &81.5  &63.1  &19.6   &2.5    &86.4\\
Z. Zhang et al \cite{ZZ}   &2    &  46       &88.9  &99.0  &79.1   &36.0  &85.5  &92.8  &93.8   &13.7   &86.7 \\
C. Herry et al \cite{HFSW}  &1    &6          &83.13  &98.93  &81.14   &31.93  &77.50  &79.05  &83.25   &6.91   &82.70 \\
S. Chen et al \cite{C}    &1      &33       &98.4  &95.4  &29.5   &38.4  &70.8  &85.1  &-      &-      &93.1 \\
U. Acharya et al \cite{A}    &1      &-       &91.64  &85.17  &89.04   &94.76  &95.08  &95.21  &94.69      &93.47      &93.1 \\
S. Mathews et al \cite{MKB}    &1      &26       &-  &-  &88.39   &33.63  &77.74  &69.2  &-      &-      &- \\\midrule
Proposed  method         &   1& 19                         &85.56 &98.94 &80.37  &26.9  &80.34 &88.91 &82.80  &10.45  &85.02\\
                                          &   2& 38                         &86.67 &98.97 &80.92  &27.02 &80.17 &90.32 &81.22  &12.09  &86.02\\  \bottomrule
\end{tabular}}
\end{table*}

\subsection{Performance comparison}
A comparison of the proposed algorithm with the state-of-the-art algorithms is shown in Table \ref{Table:Comparision}. All results shown in Table \ref{Table:Comparision} are trained and validated on the MIT-BIH arrhythmia database. Moreover, except \cite{A}, all studies comply with the AAMI standard and follow DS1 and DS2 division schemes proposed in \cite{DMR} as we do.

The first six methods in Table \ref{Table:Comparision} are traditional methods, and the latter two are deep learning approaches.
The methods proposed in \cite{C} and \cite{HFSW} depend on only a single ECG lead, while other methods \cite{DMR,LM1,YBM}, and \cite{ZZ} depend on two-lead ECG signals.
In the traditional methods, for the feature extraction, \cite{DMR} and \cite{LM1} consider RR interval, ECG landmark features such as QRS duration, T-wave duration, P-wave flag, 2D vectocardiogram (VCG) loop and others, heartbeat segmentation information of P, QRS, T-wave onset, offset points and so on, as their features, whereas \cite{YBM} and \cite{ZZ} consider a combination of landmark and dynamic features, wavelets, ICA and RR interval as their features. The phase information determined by the synchrosqueezing transform is taken as a feature in \cite{HFSW}.
The feature selection is conducted in \cite{ZZ} to determine the best features. A random projection was considered in \cite{C} to determine the final features. Then, \cite{DMR} and \cite{LM1} choose LD as their classifiers, while the others choose SVM. For the deep learning approaches, \cite{MKB} used the RBM and the DBN after extracting feature sets of RR intervals, heart-beat intervals, and segmented morphology, while \cite{A} developed a 9-layer DNN for the ending-to-ending learning.

As observed in Table \ref{Table:Comparision}, the proposed classification model achieves a balanced classification rate. The sensitivity of all beat classes is over 80\% and the overall accuracy is also greater than 85\%.
The overall accuracy of \cite{C} was up to 93.1\%, which at first glance is better than ours, but the authors neglect classes of small sizes; specifically, the heartbeats in the F class are not considered in the analysis, and the overall accuracy of the S class is not ideal. As the purpose of the proposed automatic heartbeat classification is detecting various kinds of non-lethal arrhythmia, we consider our result better.
\cite{A} achieved a 93.47\% accuracy, in which synthetic data is used to overcome the classes imbalance. Besides, its experimental scheme is not inter-patient CV, so its performance in the inter-patient CV scheme is not clear.

The results reported in \cite{YBM} and \cite{ZZ} are over 86\%, which are better than our performance. This is expected since two leads and more features (\cite{YBM} uses 132 initial features and \cite{ZZ} uses 46 features) are used in these papers.
However, the sensitivity of the F class is only 19.6\%, with the positive predictivity $2.5$\% in \cite{YBM}. The result in \cite{ZZ} is more balanced and closer to ours, but compared with our approach, they use a complicated feature ranking approach for selecting features and dimension reduction.
The purpose of \cite{MKB} is to classify two abnormal heartbeats of S class and V class. Although the performance of S class and V class is excellent, the overall accuracy from the confusion matrix provided in \cite{MKB} is below 80\%. We thus consider our approach superior.

In order to compare the multi-lead methods from Table \ref{Table:Comparision}, we also show the results from training the classifier using the merged decision by the features described in Section 5. The features are extracted form both channels of the aforementioned databases, and the results are shown in Table \ref{Table:Comparision}. The overall accuracy is greater than 86\% and the sensitivity of all beat classes is over 80\%.

In conclusion, among the compared algorithms, the proposed method yields comparable performance, in terms of overall accuracy, as well as the V, S and F classes.

\subsection{Imbalanced dataset}
The imbalanced dataset issue is challenging to all automatic heartbeat classification methods. As considered in \cite{LM1}, perhaps the division of DS1 and DS2 is not suitable for certain beat classes. For example, the atrial premature beats of one record in DS2 account for more than half of the S class. In this work, the weighted kernel SVM is considered, and an extra database is taken into account to alleviate the imbalanced dataset issue.

\subsection{Real-time implementation}
The proposed classifier is trained on a computer with 16 GB RAM and 2.71 GHz Intel core i5 processor. The algorithm is developed in MATLAB R2016a. It takes approximately 81.83 seconds to complete one training process. Once the training of ECG signals is complete, the classification of ECG heartbeat is fast. The critical step in the proposed classification system is the feature extraction based on the AFD. The computation complexity of the AFD algorithm is $\mathcal{O}(N\log N)$ \cite{GKQW} and the computation time of the AFD with $N=10$ is $0.189$s. Note that this time is much shorter than the time needed to finish one heartbeat, even when the heart rate is as fast as 180 beats per minute. Thus, the proposed algorithm has the potential for a real-time monitoring system. Real-time implementation will be the focus of future work.
Since none of the traditional methods shown in Table \ref{Table:Comparision} provide the training time, a comparison cannot be made. Deep learning requires a long training time and a specialized hardware, such as GPU, to efficiently train the algorithm. For example, the CNN in \cite{A} is trained with two Intel Xeon 2.40 GHz (E5620) processors and 24GB of RAM, which takes approximately 9,573.2 seconds to complete one training epoch.

\subsection{Limitations and Future work}
The proposed approach has some limitations. The data is from a publicly available database that is not collected from equipment tailored for telemedicine. Also, the database size is limited. Therefore, a larger database collected from professional mobile devices for the telemedicine is needed to confirm the practical performance of the proposed algorithm.

We focus only on the ECG signal in this paper, while the proposed algorithm has the potential to be applied to other biomedical signals,  which will be explored in future work. Furthermore, the possibility of extracting features from AFD to train the deep learning framework to improve the overall performance will be considered in the future study.

\section{Conclusion}\label{Section:Conclusion}
This paper presents a novel automatic heartbeat classification based on the recently developed signal processing tool, AFD, which is applied to capture morphological characteristics of the ECG signal from a complex analysis perspective. The SVM classifier is considered for automatic classification, with input features including AFD-derived IFs, amplitude at R-peak location, QRS duration, and three RR interval features.
The heartbeat classification performance on the MIT-BIH arrhythmia database is compatible with other state-of-the-art methods that depend on a huge number of features and/or multi-lead ECG. This method has the potential to be used in an ambulatory ECG monitoring device or a mobile health device for real-time diagnosis of  non-life threatening arrhythmias.

\section*{Acknowledgment}
This study is supported by the research grants:The Science and Technology Development Fund of Macao SAR FDCT 079/2016/A2, 0123/2018/A3, and MYRG2017-00218-FST.

\ifCLASSOPTIONcaptionsoff
  \newpage
\fi

\end{document}